\documentstyle[aps,prl,twocolumn,epsf]{revtex}
\begin{document}
\twocolumn[\hsize\textwidth\columnwidth\hsize
\csname @twocolumnfalse\endcsname

\title{Theory of Coherent $c$-Axis Josephson 
Tunneling between Layered Superconductors}

\author{ G. B. Arnold }
\address{ Department of
  Physics, University of Notre Dame, Notre Dame, IN 46556}
\author{ R. A. Klemm}
\address{ Materials Science Division, Argonne National Laboratory, Argonne, IL 60439-4845}

\maketitle

\begin{abstract}

We calculate exactly the Josephson current for $c$-axis coherent tunneling between two layered superconductors, each with internal coherent tight-binding intra- and interlayer quasiparticle dispersions. Our results also apply when one or both of the superconductors is a bulk material, and include the usually neglected effects of surface states.  For weak tunneling, our results reduce to our previous results derived using the tunneling Hamiltonian.   Our results are also correct for strong tunneling. However, the $c$-axis tunneling expressions of Tanaka and Kashiwaya are shown to be incorrect in any limit. In addition, we consider the $c$-axis coherent critical current between two identical layered superconductors twisted an angle $\phi_0$ about the $c$-axis with respect to each other.  Regardless of the order parameter symmetry, our coherent tunneling results using a tight-binding intralayer quasiparticle dispersion are inconsistent with the recent $c$-axis twist bicrystal Bi$_2$Sr$_2$CaCu$_2$O$_{8+\delta}$ twist junction experiments of Li {\it et al.} 
\end{abstract}
]
\narrowtext

%\newpage
\section{Introduction}

One of the most interesting features of superconductivity is the Josephson effect.  This occurs when flat surfaces of two superconductors are brought together, forming a uniform junction.  A supercurrent flows without a voltage drop across the junction, provided that the properties of the superconductivity in the two superconductors are compatible.  For conventional superconductors in which both the normal state and superconducting properties are isotropic, it doesn't matter which particular crystal surfaces are employed to form the junction.  

However, for anisotropic superconductors, the junction orientation can be very important.  Even if the superconducting order parameter (OP) is isotropic, the intrinsic anisotropic normal state properties of a layered superconductor make the properties of Josephson junctions involving one or more layered superconductors different from those formed from two isotropic materials.  For example, Josephson junctions between an isotropic, conventional superconductor and a layered superconductor can differ greatly, depending upon whether the junction is on the top or an edge of the layered superconductor. This is especially true if the layered superconductor has an anisotropic OP.  In addition, the properties of Josephson junctions formed between two layered superconductors depend strongly upon the junction orientation, especially if the OPs are anisotropic. 

 In this paper, we consider the case of a Josephson junction formed between two layered superconductors stacked on top of each other along the $c$-axis.  Our results can be easily modified to include the related problems in which one or both of the superconductors is bulk, rather than layered.  By treating the two superconductors as layered, the surface states that form near the junction appear naturally in the calculation.  These surface states affect the Josephson current results, even when the limit of tunneling between two isotropic superconductors is taken.  With one exception, these surface states have previously been neglected.  In addition, when the quasiparticle dispersions of the two superconductors are different, such as for different materials or identical tight-binding materials that are rotated with respect to one another, an impedance mismatch occurs, reducing the coherent critical current.

\begin{figure}[htb]
\vspace*{-1.0cm}
\epsfxsize=1.0cm
\centerline{\hspace*{-7cm}\epsffile{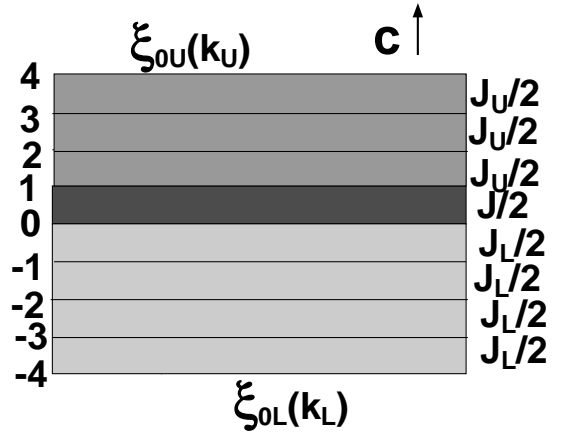}}
\vspace*{6cm}
\caption{Sketch of the junction under study.  Layer numbers and interlayer tunneling strengths are indicated, along with the general forms for the bare quasiparticle dispersions.}\label{Fig1}
\end{figure}

We consider the Josephson tunneling along the $c$-axis between two distinct layered superconductors, assuming all of the tunneling processes can be taken to be purely coherent, as pictured in Fig. 1.  In Section II, we solve for the Green's function in each layer, keeping the quasiparticle dispersions and OP symmetries fully general. The effects of surface states are explicitly included in the calculations, and the various tunneling strengths are included exactly.  In Section III, we derive the temperature ($T$) dependence of the superconducting gaps under three specific assumptions about the quasiparticle band structures and three different OP symmetries in each layered superconductor.  These quasiparticle dispersions are chosen to model the experimental cases of underdoped, optimally doped, and overdoped Bi$_2$Sr$_2$CaCu$_2$O$_{8_+\delta}$, (BSCCO), respectively, with respective Fermi surfaces that we denote as FS1, FS2, and FS3.   However, the dispersions for overdoped samples can also be made to fit the three-dimensional free-quasiparticle dispersions of conventional bulk superconductors.  The OP symmetries chosen are the isotropic $s$-wave and the tight-binding $d_{x^2-y^2}$-wave and ``extended-$s$''-wave OP functions, respectively.  In Section IV, we solve for the gaps for the symmetric case of two identical superconductors. Then, in Section V, we calculate the Josephson tunneling current across the two layered superconductors, assuming that the coupling across the junction is not stronger than the intrinsic coupling within each layered superconductor.  Our final results for the tunneling current can be used to calculate the $c$-axis Josephson critical current $I_c$, and can be generalized to the standard case of two bulk, conventional superconductors, for which the usually neglected surface states are explicitly included.  Then, in Section VI, we present detailed $I_c(\phi_0)$ results for coherent $c$-axis twist junctions between identical coherent layered superconductors.  Finally, we discuss our results in Section VII.

\section{Model and Procedure for Solving it}

We assume that two layered superconductors are placed with the upper ($U$) one on top of the lower ($L$), and that the contact between them is sufficiently strong that quasiparticle tunneling between them occurs.  For simplicity, we assume each superconductor consists of $N\gg1$ layers separated an equal distance $s$ apart. We index the layers with $n, m$, where $-N+1\le n, m\le N$.  In the $U$ half space, $1\le n, m\le N$, and in the $L$ half space, $-N+1\le n, m\le0$, as pictured in Fig. 1.  Within each layer in the $\eta=L,U$ half space, the quasiparticles propagate with dispersion $\xi_{0\eta}({\bf k})$ and have the gap function $\Delta_{\eta}({\bf k})$, where ${\bf k}=(k_x,k_y)$ is a two-dimensional wavevector.  We assume $\Delta_{\eta}$ is independent of layer index.  This assumption will be checked in Section III, and is found to be usually valid.   Between adjacent layers in each half space, the quasiparticles tunnel with matrix element $J_{\eta}/2$.  At the junction between layers 0 and 1 where the $L$ and $U$ half spaces meet, the quasiparticles tunnel with matrix element $J/2$.  Since we are not interested in spin-dependent effects, we assume the quasiparticles are spinless fermions, only taking account of the spin values in counting the number of quasiparticles.  We set $k_B=c=\hbar=1$.

We shall focus on the general procedure for evaluating $\hat{G}$.  The details are given in the Appendix. 
In order to calculate the Josephson current across the junction between the two half spaces, we first find the form of the finite temperature Greens' functions matrix $\hat{G}$.  This matrix is the product of two matrices, one of rank $2N$, with elements indexed by the layers $n,m$, and the other the Nambu matrix of rank 2, with elements $G$, $F$, $-G^{\dag}$, and $F^{\dag}$ in the usual cyclic order beginning with the upper left hand position.   This Nambu matrix is represented by the Pauli matrices $\tau_i$ for $i=1,2,3$, plus the rank two identity matrix $\tau_0$.  We let $\omega$ represent the Matsubara frequencies.  

We first begin by constructing  the Green's function matrix $\hat{\cal G}$ for a bulk layered superconductor.  We then add a perturbation with the particular form that decouples the Green's functions in each half space from each other.  The resulting Green's functions $\hat{g}^{\eta}$ of two single half spaces have parameters appropriate for each half space.  We then couple these two half-space Green's functions together.  

The Green's function matrix $\hat{\cal G}^{\eta}$ for a bulk superconductor of type $\eta$ satisfies
\begin{equation}
[i\omega-\xi_{0\eta}\tau_3-\Delta_{\eta}\tau_1+\hat{\cal J}^{\eta}\tau_3]\hat{\cal G}^{\eta}=\hat{1},\label{bulkmatrix}
\end{equation}
where 
the layer space matrix $\hat{\cal J}^{\eta}$ has the elements
\begin{equation}
{\cal J}_{mn}^{\eta}={{J_{\eta}}\over{2}}\Bigl(\delta_{m,n+1}+\delta_{m,n-1}\Bigr).\label{bulkJ}
\end{equation}
Terms not containing a Pauli matrix are implicitly proportional to $\tau_0$.  
In a bulk layered superconductor, it is then possible to Fourier transform this expression, as we did in the Appendix, with the result differing only from that of a bulk, three dimensional superconductor by the corrugated cylinder form of the quasiparticle dispersion, $\xi_{\eta}({\bf k}, k_z)=\xi_{0\eta}({\bf k})-J_{\eta}\cos(k_zs)$.  However, for a half space, such Fourier transformation is not permissible, and so we must keep the layer indices $n,m$ explicitly.    

To construct $\hat{g}^{\eta}$, we first remove the tunneling matrix elements across the junction between the two superconductors by adding the perturbation $\hat{V}^{\eta}\tau_3$, with elements
\begin{equation}
V_{mn}^{\eta}=-{{J_{\eta}}\over{2}}(\delta_{m0}\delta_{n1}+\delta_{m1}\delta_{n0}),\label{veeeta}
\end{equation}
taking account of the restrictions $n,m\le0$ for $\eta=L$ and $n,m\ge1$ for $\eta=U$.
The perturbation $\hat{V}^{\eta}$ ``cuts'' the bonds connecting the two identical
half spaces which terminate at layers 0 and 1. \cite{KS} We then solve for the elements $g_{mn}^{\eta}$ of $\hat{g}^{\eta}$ satisfying
\begin{equation}
\hat{g}^{\eta}=\hat{\cal G}^{\eta}+\hat{\cal G}^{\eta}\hat{V}^{\eta}\tau_3\hat{g}^{\eta}.\label{geeeta}
\end{equation}
Solutions for each $\hat{g}^{\eta}$ are given in the Appendix.

So far, we have found the expressions for two distinct half-space Green's functions, which are electronically uncoupled from each other, since no quasiparticle propagation from one half space to the other has yet been introduced. We thus couple $\hat{g}^{U}$ and $\hat{g}^{L}$ together via the local
perturbation $\hat{\cal J}$ with matrix elements
\begin{equation} {\cal J}_{mn}=
{{J}\over{2}}(\delta_{m0}\delta_{n1}+\delta_{m1}\delta_{n0}).\label{coupling}
\end{equation}
For tunneling strength comparisons, we then let
\begin{equation} \gamma={{J^2}\over{J_L J_U}}.\end{equation}
The exact solution to this problem of coupled half spaces then yields the full Green's function matrix $\hat{G}$, with matrix elements $G_{mn}$, which is given in the Appendix.

\section{Gap Equation}
In order to obtain the temperature dependence of the Josephson critical
 current,
we need to solve a gap equation for the temperature dependence of the gap.  
For
this we make a simple assumption of a BCS-like equation
\begin{equation} \Delta_{\eta n}({\bf k},T)={T\over2}\sum_{\omega}\sum_{{\bf k}'}
\lambda_{\eta}({\bf k},{\bf k}')
{\rm Tr}[(\tau_1+i\tau_2)G_{nn}({\bf k}',\omega)].\label{gapeqn}
\end{equation} This
layer-dependent gap function is in contradiction to the layer-independent gap function used
to calculate the Green's functions.  This must therefore be
regarded as the first correction to the assumed constant gap,
which determines the Green's function on the right hand side of
this equation. Ideally, this should be a small correction, if our
initial assumption is justifiable.

In the Appendix, we found an expression for $G_{00}{{J_L}\over{2}}\tau_3$. Using this result, letting 
\begin{equation}
\Xi_{\eta}={{2}\over{\xi_{0\eta}+i\Omega_{\eta}+[(\xi_{0\eta}+i\Omega_{\eta})^2-J_{\eta}^2]^{1/2}}},\label{xi}
\end{equation} 
where $\Omega_{\eta}=\sqrt{\omega^2+|\Delta_{\eta}|^2}$, and defining $\Xi_{\eta}'=\Re(\Xi_{\eta})$ and $\Xi_{\eta}''=\Im(\Xi_{\eta})$, 
the equation for the gap at the interface in the lower superconductor
is 
\begin{equation}
\Delta_{L0}({\bf k},T)=T\sum_{\omega}\sum_{{\bf
k}'}{{\lambda_L({\bf k},{\bf k}')}\over{D({\bf k}',\omega)}} f({\bf k}',\omega),\label{exactgap}
\end{equation}
where
\begin{equation}
f={{\Delta_L\Xi_L''}\over{\Omega_L}}+{{\Xi_U''|\Xi_L|^2J^2\Delta_U}\over{4\Omega_U}},
\end{equation}

\begin{eqnarray}
D&=&
{J^2\over2}\Bigl[\Bigl({{\omega^2+
\Delta_L\Delta_U\cos(\phi_L-\phi_U)}\over{\Omega_U\Omega_L}}\Bigr) \Xi_U''\Xi_L''\cr
& &\cr
& &\qquad
-\Xi_U'\Xi_L'\Bigr] +1+{J^4\over{16}}|\Xi_U|^2|\Xi_L|^2,
\label{D}
\end{eqnarray}
and $\phi_L-\phi_U$ is the phase difference of the OPs across the junction.

As a quick check
of this result, we take $L=U$ and $\gamma=1$, and find that this reduces to
\begin{eqnarray}
\Delta_L({\bf k},T)& =&T\sum_{\omega}\sum_{{\bf k}'}\lambda_L({\bf
k},{\bf k}'){{\Delta_L({\bf
k}',\omega)}\over{\Omega_L({\bf k}',\omega)}}\times\cr
& &\cr
& &\times\Im\biggl[\Bigl([\xi_{0L}({\bf k}')+i\Omega_L({\bf k}',\omega)]^2-J_L^2\Bigr)^{-1/2}\biggr],\label{bulkgap}
\end{eqnarray}
in agreement with the result for a homogeneous coherent layered superconductor.

Note that in Eq. (\ref{exactgap}), the proximity effect couples the 
gap function at the interface in $L$ to the gap function in $U$.
This disappears in the limit $\gamma\ll1$, wherein
$\Delta_0({\bf k},T)$ becomes the gap function at the surface of
$L$:
\begin{eqnarray} 
\Delta({\bf k},T)_{L,surface}& =&T\sum_{\omega}\sum_{{\bf k}'}\lambda_L({\bf
k},{\bf k}')\times\cr
& &\cr
& &\times{{\Delta_L({\bf
k}',\omega)\Xi_L''({\bf k}',\omega)}\over{\Omega_L({\bf k}',\omega)}}, \label{surfacegap}
\end{eqnarray}
in agreement with previous results. \cite{LK}

The gap functions which go into the right hand side of
(\ref{exactgap}) are the spatially constant (``zeroth order'')
bulk gap functions, obtained from (\ref{bulkgap}) for $L$, and an
analogous equation for $U$.  We use these to calculate
$\Delta_{\eta}({\bf k}, T)$ and $\Delta({\bf k}, T)_{\eta,surface}$ for $\eta=L,U$, and
compare the magnitudes of these to the bulk results. Ideally, the
difference in magnitudes between the bulk gaps and the interface
gaps should be small.  In the following section, we will
investigate this for the special case of a symmetric junction.

\section{Gaps for the Symmetric Case}

In this section, we assume that $L=U$ and solve for the bulk gap,
and the gap at the interface with $\gamma\ne 1$.
To find the bulk gap, we fix $T_c$ at 9 meV, and solve
Eq. (\ref{bulkgap}) for the temperature dependent gap.  We assume each superconductor has a single OP, which for simplicity, we limit to $s$, $d_{x^2-y^2}$, or extended-$s$-wave symmetry. We index these OPs by $\zeta=s,d,es$. For each OP, we take the pairing interaction on the $\eta=L,U$ superconductor to have the form
\begin{equation}
\lambda_{\zeta\eta}({\bf k},{\bf k}')=\lambda_{\zeta\eta}\Psi_{\zeta}({\bf k})\Psi_{\zeta}({\bf k}'),\label{pairing}
\end{equation}
where $\Psi_s({\bf k})=1$, $\Psi_d({\bf k})= \cos(k_xa)-\cos(k_ya)$, and $\Psi_{es}({\bf k})=\{[\cos(k_xa)-\cos(k_ya)]^2+\epsilon^2\}^{1/2}$, where $\epsilon\ll1$.

\begin{figure}[htb]
\vspace*{-0.0cm}
\epsfxsize=8.0cm
\centerline{\epsffile{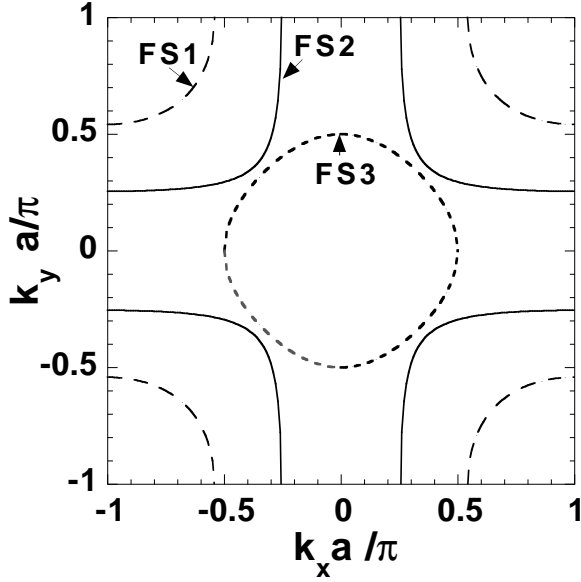}}
\caption{Fermi surfaces studied for tunneling between two layered superconductors.  Dashed curves:  underdoped FS1.  Solid: optimally doped  FS2.  Dotted:  overdoped FS3.}\label{Fig2}
\end{figure}

By fixing $T_c$, we determine the value of $\lambda_{\zeta\eta}$. 
These then determine the $T$ dependence of
the s-wave or d-wave gaps, respectively, via Eq. (\ref{bulkgap}), and the extended-s-wave gap with $\epsilon\rightarrow0$ has an identical $T$ dependence to that of the d-wave gap.  We
take the in-plane dispersion to be
\begin{eqnarray} 
\xi_{\eta0}({\bf k})&=&J_{\eta}-J_{\eta ||}[\cos(k_xa)+\cos(k_ya)\cr
& &\cr
& &-\nu\cos(k_xa)\cos(k_ya)-\mu],
\label{dispersion}\end{eqnarray}
where the part proportional to $J_{\eta ||}$ is chosen to
approximate the in-plane dispersion relation for BSCCO. The
value of $J_{\eta ||}$ is thus taken to be 500 meV.  We choose three sets of parameters $\nu$ and $\mu$, which determine the details of the quasiparticle dispersion, and the resulting shape of the two-dimensional Fermi surface, pictured in Fig. 2. For a heavily underdoped sample, we choose $\mu=-1.3$, $\nu=1.3$, with Fermi surface denoted FS1 in Fig. 2.  For the tight-binding dispersion appropriate for an optimally doped sample of BSCCO, we take $\nu=1.3$ and $\mu=0.6$. This dispersion has the Fermi surface denoted FS2 in Fig. 2. In addition, for a heavily overdoped sample, we choose $\nu=0$ and $\mu=1.0$, with the Fermi surface FS3 in Fig. 2. The only
remaining free parameter is the value of $J_{\eta}/2$, the interlayer
overlap integral. We therefore perform our calculations for
three different values: $J_{\eta}=25, 50, 100$ meV.  In each case, $J_{\eta}=25$ meV gives results which closely approximate the weak hopping limit $J_{\eta}\rightarrow0$, as has been verified by explicitly checking our results with $J_{\eta}=1$ meV.  However, for that small an interlayer hopping parameter, one needs to use a much finer grid for the Brillouin zone integration, in order to obtain sufficient accuracy.

For each OP symmetry $\zeta$, we calculate $\Delta_{\zeta}(T)$ from the symmetric gap equation.  We note that $\Delta_{es}(T)=\Delta_d(T)$ for $T_{cd}=T_{ces}$, which occurs for $\lambda_{es}=\lambda_d$.  In the case $\gamma=1$, the combined junction of two layered superconductors is just the same as a single layered superconductor, as long as the two halves are not twisted with respect to each other.  However, for $\gamma<1$, the central junction is different than the intrinsic ones in each layered superconductor.  In this case, surface states can arise, and the gap $\Delta_{\mu}$ can in principle depend upon the layer index.  This would be particularly true in the case of two incompatible OP components, which has been considered in detail for a cylindrical Fermi surface previously. \cite{KRS}

\begin{figure}[htb]
\vspace*{-0.0cm}
\epsfxsize=6.0cm
\centerline{\epsffile{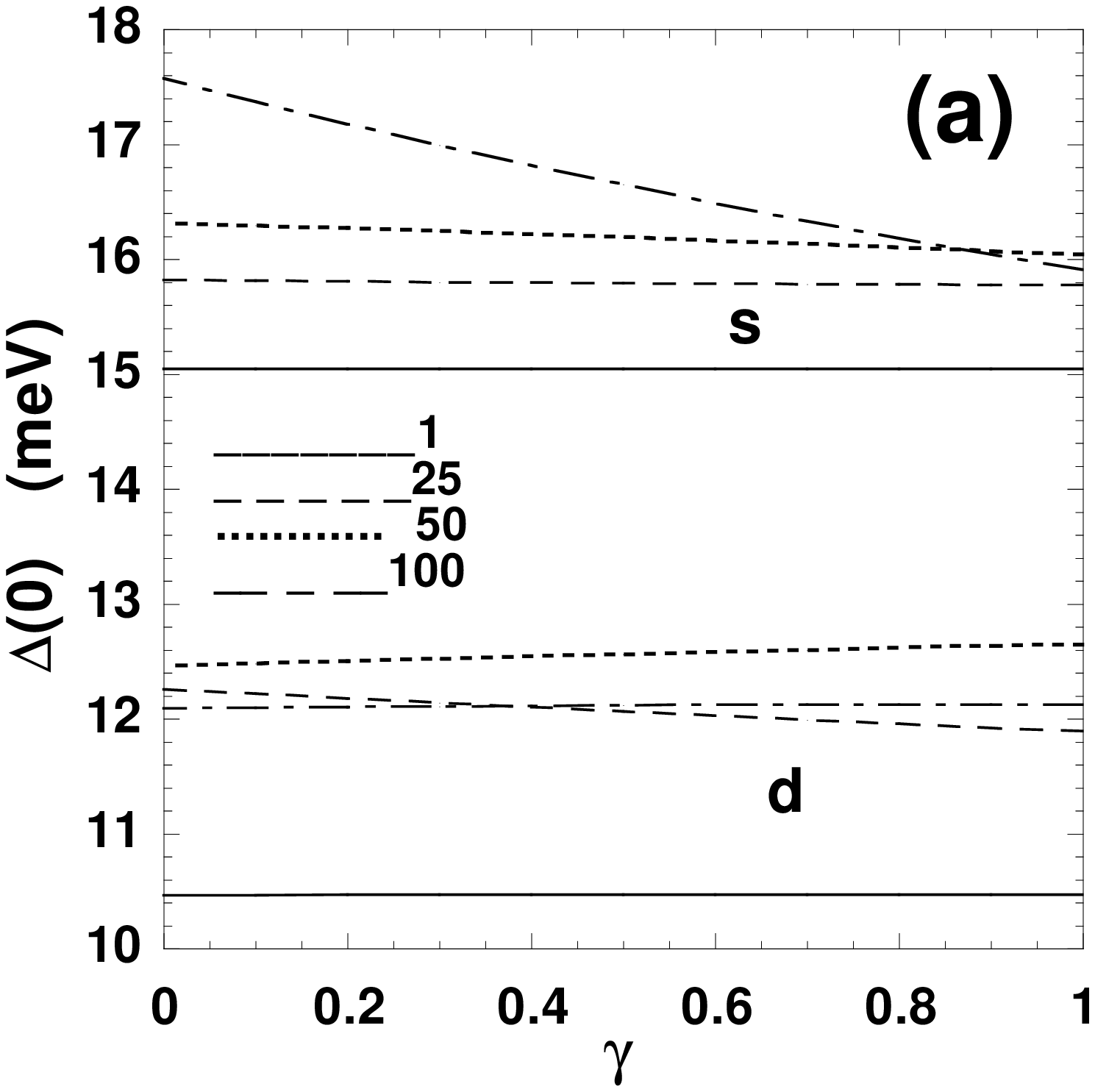}}
\end{figure}
\begin{figure}[htb]
\vspace*{-0.8cm}
\epsfxsize=6.0cm
\centerline{\epsffile{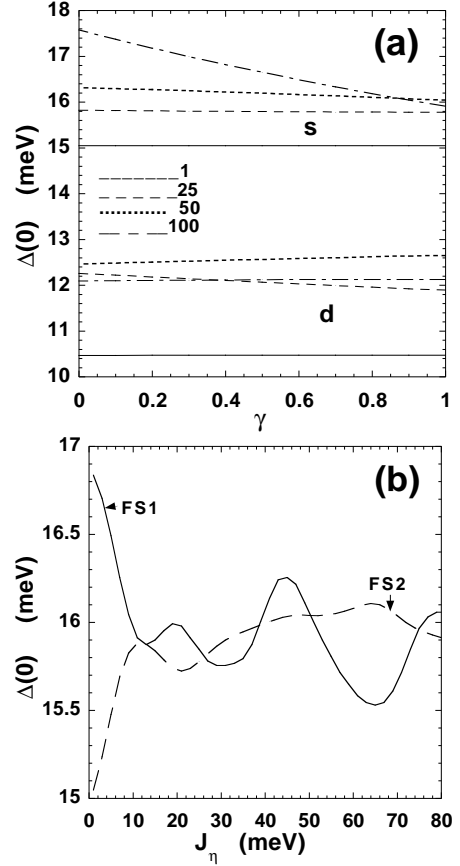}}
\vspace*{0.3cm}
\caption{(a) $s$- and $d$-wave superconducting gap magnitudes at $T=0$ as functions of $\gamma$ for FS2 with $J_{\eta}=25, 50, 100$ meV. (b) Plots of the $T=0$ $s$-wave gap magnitudes at $\gamma\ll1$ for Fermi surfaces FS1 and FS2,  as a function of $J$ in meV.}\label{Fig3}
\end{figure}

For the Fermi surface FS2 and only one OP component, if $\gamma\ll1$ and $J_{\eta}=0$, then each layer would be completely isolated from each other, and the gap wouldn't depend upon the layer index, reducing to the two-dimensional mean-field value of independent layers.  Hence, the question of whether the gap depends upon the layer index can be more easily answered by checking whether the gap depends upon the intrinsic interlayer hopping strength $J_{\eta}$ and on the impedance mismatch parameter $\gamma$.  In Fig. 3a, we  plotted $\Delta_s(0)$ and $\Delta_d(0)$ for $J_{\eta}=25,50, 100$ meV, as functions of $\gamma$ ($0\le\gamma\le1$), for the optimally doped Fermi surface FS2, evaluated with $\nu=1.3$ and $\mu=0.6$, pictured in Fig. 2.  Although there is a weak $\gamma$ dependence of $\Delta_d(0)$ for $J_{\eta}=100$ meV, 
all other cases have essentially no significant $\gamma$ dependence, so that the gap values are essentially the same as for independent layers.  The slightly greater $\gamma$ dependence of $\Delta_d(0)$ for $J_{\eta}=100$ meV apparently arises because the Fermi energy $E_F$ is not much further than $J_{\eta}$ from the top of the quasiparticle band. In Ref. (3), it was also found that for a free particle dispersion within the layers that the gap at the surface $(\gamma=0$) was the {\it same} as that in the bulk ($\gamma=1$).  Thus, our results for a finite in-plane bandwidth closely approximate that infinite bandwidth limit.

We also studied $\Delta_{s}(0)$ as a function of $J_{\eta}$ at $\gamma\ll1$ for Fermi surfaces FS1 and FS2, and plotted the results in Fig. 3b.  Again, we found very little $J_{\eta}$ dependence to the $s$-wave gap magnitudes for $\gamma\ll1$, so that it is generally safe to take the gap to be the bulk value calculated far from the central junction location. We also found that $\Delta_{\zeta}(T)/\Delta_{\zeta}(0)$ for $\zeta=s,d$ with Fermi surface FS2 were rather independent of $J_{\eta}$ for $J_{\eta}=1,25,50,100$ meV, each of the curves differing only slightly from the ordinary BCS curves for $\Delta(T)$.

\section{Josephson Tunneling Current}
The tunneling current across the junction is given by
\begin{equation} I=-ieJ\sum_{{\bf k}}\langle
c_0^{\dagger}({\bf k})c_1({\bf k})-c_1^{\dagger}({\bf k})c_0({\bf k})\rangle 
\end{equation}
where the $c_i({\bf k}),c_i^{\dag}({\bf k})$ are creation and
annihilation operators for electrons  with
wavevector ${\bf k}$ in the $i^{th}$ layer.  The angular brackets indicate a thermodynamic
equilibrium average. In terms of the full space Green's functions, this is
\begin{equation} I=ieJT\sum_{{\bf k}}\sum_{\omega}{\rm Tr}[{1\over2}(\tau_0+\tau_3)
(G_{10}-G_{01})],\end{equation}
where we have suppressed the 
$({\bf k}, \omega)$ dependence of the Green's functions.
 
In the Appendix, the matrices $G_{01}$ and $G_{10}$ are given, and the trace evaluated.  We find

\begin{equation}
I=eT\sum_{\omega}\sum_{\bf k}{{ J^2\Delta_L\Delta_U\Xi_L''\Xi_U''}\over{\Omega_L\Omega_UD}}
\sin(\phi_L-\phi_U),\label{exactcurrent}
\end{equation}
where $\Xi_{\eta}$ is given by Eq. (\ref{xi}) and $D$ is given by Eq. (\ref{D}).

The equations derived above relate the supercurrent through the
interface between two layered superconductors to the phase
difference across this interface.  We can find the critical
current $I_c^J$  of this interface by varying the phase difference until
the maximum current is obtained. We can vary the value of
$\gamma$, which we term ``the impedance match parameter'', to
obtain $I_c(\gamma)$.  We can also vary $T$ to obtain $I_c(T)$.  Finally, we can investigate
all of the above for $s$-, and $d$-, and extended-$s$-wave gaps, and mixtures
thereof.

To order $J^2$, we can set $D\rightarrow 1$, and $I_c^J$ obtained from Eq. (\ref{exactcurrent}) reduces precisely to our previous result, Eq. (8) of Ref. (4), derived using the tunneling Hamiltonian, provided only that one neglects the incoherent tunneling, and replaces $J$ with $2{\cal T}_0$ in the coherent tunneling part. \cite{KARS} However, this result differs greatly from that found by Tanaka and Kashiwaya (TK). \cite{TK}  TK did not correctly derive the coherent $c$-axis tunneling between two layered superconductors, but instead attempted to approximate the tight-binding quasiparticle dispersion in the $c$-axis direction by treating the corrugated Fermi surface as a narrow belt around a spherical Fermi surface. This procedure leads to the uncontrolled approximation of dividing zero by zero.  Thus, Eq. (100) of TK, obtained in the weak tunneling limit of their calculation, is not correct for coherent tunneling.  Instead, it happens (from compensating errors) to be correct for purely incoherent tunneling, as in the model of Ambegaokar-Baratoff (AB). \cite{AB}  
However, the subsequent Eq. (101) of TK is still not quite correct for any type of tunneling, and Eqs. (102)-(104) of TK are completely wrong for both incoherent and coherent tunneling. A modified version of Eq. (101), correct for incoherent tunneling between a conventional and an unconventional superconductor, was used appropriately in fits to $c$-axis tunneling between Pb and BSCCO. \cite{MK}

In this notation, we can also investigate the case of the coherent tunneling matrix elements depending upon the in-plane wavevectors, $J_{\eta}({\bf k}_{\eta}, {\bf k}_{\eta})=J_{\eta}\varphi^2({\bf k}_{\eta})$, and $J({\bf k}_L, {\bf k}_U)=J\varphi({\bf k}_L)\varphi({\bf k}_U)$, where $\varphi({\bf k})=|\cos (k_xa)-\cos(k_ya)|$, as suggested from band structure calculations of Liechtenstein {\it et al}. \cite{Andersen}  This form, along with the wavevector independent model, is useful for studying the coherent critical currents across a $c$-axis twist junction.  We note that $\gamma({\bf k}_L, {\bf k}_U)=\gamma$ is independent of the ${\bf k}_{\eta}$.

As long as $\gamma\le1$, the overall critical current $I_c$ will be given by the above $I_c^J$.  For both strong $\gamma=1$ and  weak $\gamma\ll1$ coupling with Fermi surface FS2 and $J_{\eta}= 1,25, 50, 1000$ meV, we calculated $I_c(T)$, normalized to its $T=0$ value, for each of the OP symmetries considered. For $\gamma\ll1$, the $s$-wave $I_c(T)/I_c(0)$ curves are almost indistinguishable from the standard AB curve, as for Fig. 2 of Ref. (4).   The other results are plotted in Fig. 4.  Note that the $d$-wave and extended-$s$-wave curves are identical, but they differ substantially from the $s$-wave results. However, as shown in  Fig. 4a, the $d$-wave (and extended-$s$-wave) $I_c(T)/I_c(0)$ curves are strongly dependent upon $J_{\eta}$, in a non-monotonic fashion, unlike Fig. (3) of Ref. (4).  Those curves in Ref. (4) were evaluated using the free-particle quasiparticle dispersion within the layers. In addition, in Figs. 4b and 4c, we plotted $I_c(T)/I_c(0)$ for the strong coupling case $\gamma=1$ for $\zeta=s, d$. respectively.  In this case, the non-monotonicity of the $d$-wave (and extended-$s$-wave) case is  similar to that in the $\gamma\ll1$ limit, but the $J_{\eta}$ variation of $I_c(T)/I_c(0)$ is stronger in Fig. 4c for $\gamma=1$ than for $\gamma\ll1$ in Fig. 4a.  Similarly, a much stronger $J_{\eta}$ variation of $I_c(T)/I_c(0)$ is seen in Fig. 4b for $\gamma=1$ than for the (not pictured) BCS-like results obtained with $\gamma\ll1$.  

\begin{figure}[htb]
\vspace*{-0.0cm}
\epsfxsize=6.0cm
\centerline{\epsffile{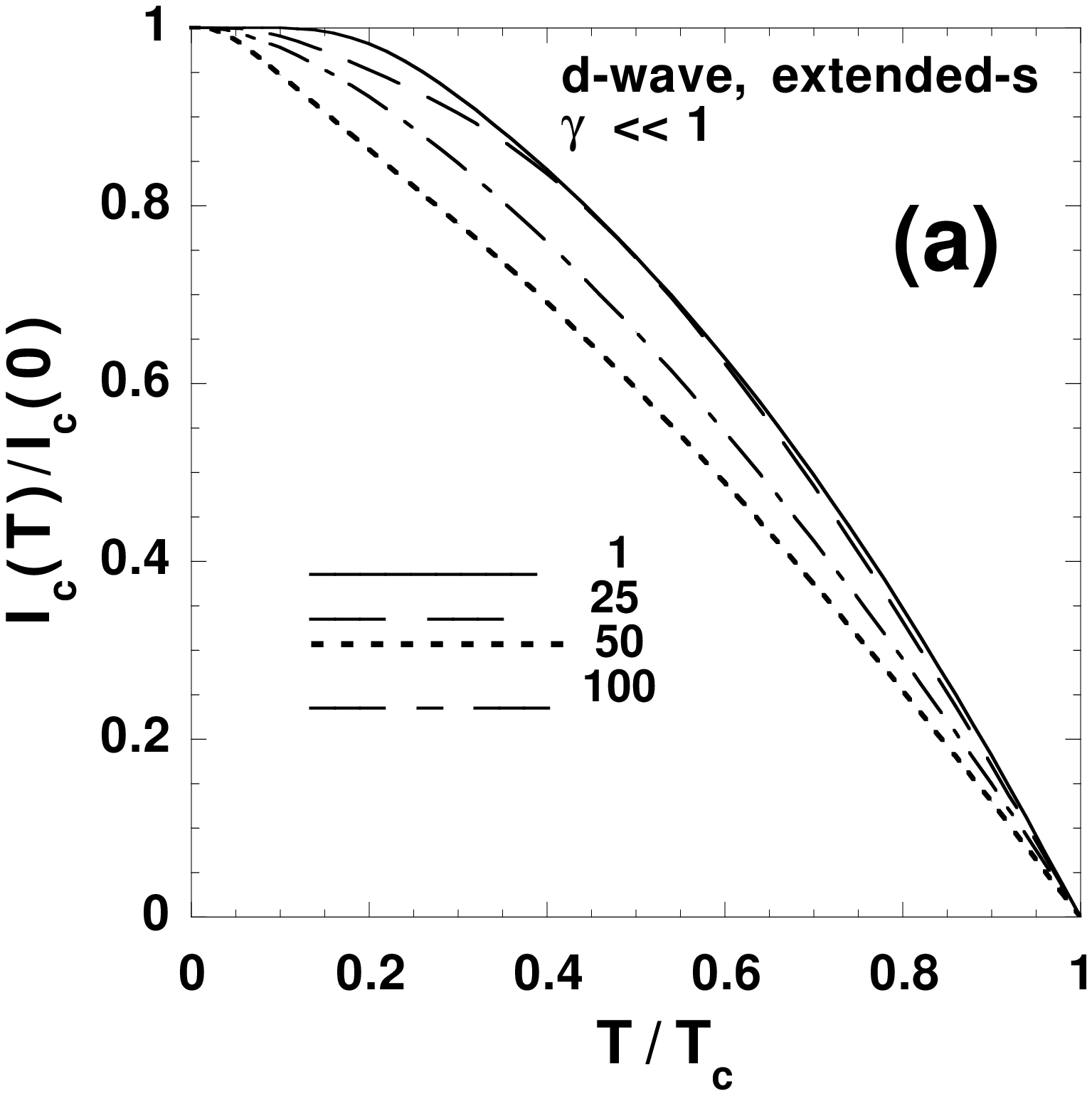}}
\end{figure}
\begin{figure}[htb]
\vspace*{-1.0cm}
\epsfxsize=6.0cm
\centerline{\epsffile{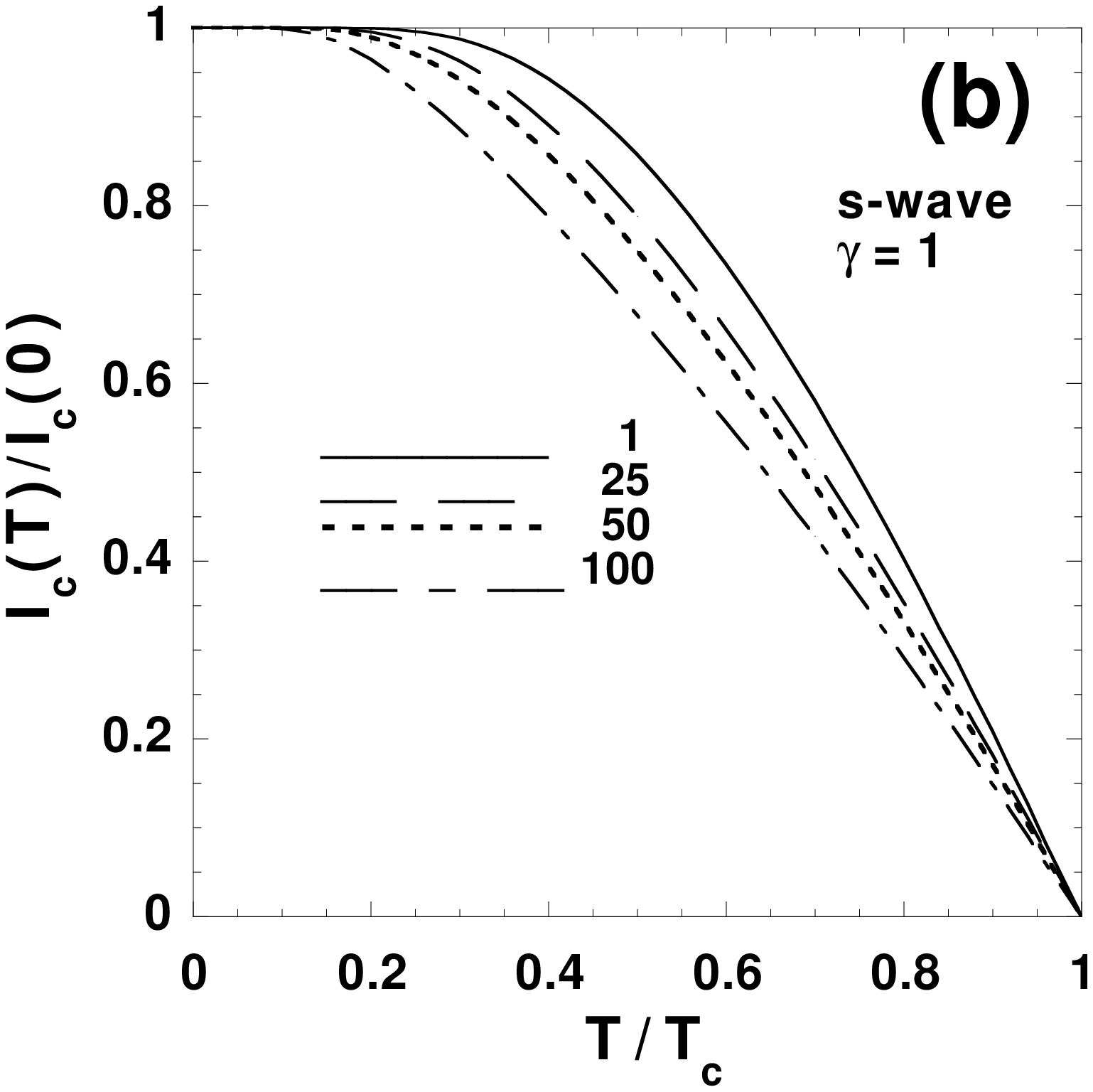}}
\end{figure}
\begin{figure}[htb]
\vspace*{-1.0cm}
\epsfxsize=6.0cm
\centerline{\epsffile{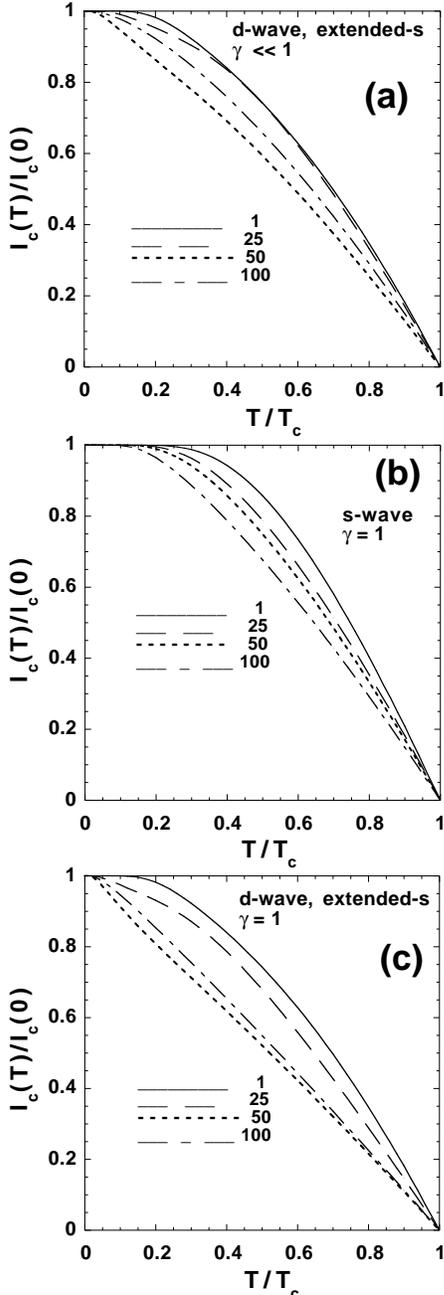}}
\caption{Plots of the normalized critical current $I_c(T)/I_c(0)$ for superconductors with  Fermi surface FS2 and  $J_{\eta}=1, 25, 50, 100$ meV. (a) $d$-wave and extended-$s$ with $\gamma\ll1$. (b) $s$-wave with $\gamma=1$.  (c) $d$-wave and extended-$s$ with $\gamma=1$.}\label{Fig4}
\end{figure}

We remark that Eq. (\ref{exactcurrent}) is a fully general expression for the $c$-axis tunneling between two layered superconductors, assuming that only one OP component exists in each superconductor, and that all tunneling processes are completely coherent.  That is, the $L$ and $R$ superconductors can have different quasiparticle dispersions (both parallel and normal to the junction), different OP symmetries, and different $T_c$ values.  Thus, Eq. (\ref{exactcurrent}) also describes the tunneling between a three-dimensional, conventional superconductor placed on the top of a layered superconductor with an unknown OP symmetry.  It also applies to the case of coherent tunneling between two three-dimensional superconductors. When one of the superconductors 
is a three-dimensional material, one simply modifies the quasiparticle dispersion $\xi_{\eta}({\bf k},k_z)=\xi_{0\eta}({\bf k})-J_{\eta}\cos(k_zs)$ far from the junction, by setting $J_{\eta ||}$ sufficiently large, and $J_{\eta}$ comparable to $J_{\eta ||}$.  Except for the case of real-space pairing with a component normal to the junction, the fact that the pairing was assumed to take place only within the layers is not important, since the layer index is dropped, so that the pairing strength is constant throughout each superconductor. Thus, a superconductor with an isotropic, three-dimensional quasiparticle dispersion would have $\xi_{\eta}({\bf k},k_z)=({\bf k}^2+k_z^2)/(2m)$ far from the junction, where $m$ is the quasiparticle effective mass.  This form is obtained from Eq. (2) by setting $J_{\eta ||}=(ma^2)^{-1}$ and $J_{\eta}=(ms^2)^{-1}$, and requiring $J_{\eta},J_{\eta ||}>>\epsilon_F$.  An anisotropic three-dimensional superconductor might have either $J_{\eta}$ or $J_{\eta ||}$ different from these values.  

We remark additionally that for the two-dimensional tight-binding Fermi surfaces, one restricts the wavevector integration to the first Brillouin zone (BZ), which for a tetragonal material has $|k_x|, |k_y|\le\pi/a$.  In the three-dimensional limit, one can use the free-particle form for the quasiparticle dispersion afr from the junction, and remove the BZ limits on the components $k_x, k_y$ of the wavevectors parallel to the junction.  Note that one still does not integrate over $k_z$, the wavevector normal to the junction, which is not a good quantum number in the presence of the junction.  Thus, even in the limit of Josephson tunneling between two conventional bulk superconductors, surface states are present, and affect the tunneling.  Except for the tunneling Hamiltonian limit $\gamma<<1$, for which we had previously included the effects of these surfaces states, \cite{KARS} our present calculation is thus the first one to correctly include the surface states in coherent Josephson tunneling between  two superconductors, whether conventional bulk, or layered.

It is important to note that when the quasiparticle dispersions in the two superconductors are not identical, there is an intrinsic impedance mismatch between the two materials.   Since in coherent tunneling, the wavevector parallel to the junction is preserved exactly, if for some wavevectors the Fermi surfaces in the two half spaces are not identical, it is not possible for the tunneling at those places in the BZ to be elastic.  That is, one cannot preserve both the momenta and the energy.  The result of this impedance mismatch is that the amplitude of the coherent tunneling, and hence the critical current, is reduced from what it would be if this impedance mismatch did not occur.  This impedance mismatch should actually occur in coherent tunneling between all inequivalent superconductors. For example, in tunneling between Nb and Pb, the Fermi surfaces, which are both three-dimensional in nature, are not identical, and some amount of coherent tunneling would be suppressed by this effect. The amount of the suppression ought to depend strongly upon the particular crystal surfaces studied at the junction location.

\section{$c$-axis twist junctions}

We now consider the case of purely coherent $c$-axis tunneling between identical layered superconductors twisted an angle $\phi_0$ about the $c$-axis with respect to each other.  This is a special case of coherent tunneling between two different layered superconductors.  However, the OP symmetry must be the same on each superconductor, with the only caveat that the OP has to arise from the local pairing interaction, Eq. (\ref{pairing}). The only difference is that the crystal orientation is rotated by $\pm\phi_0/2$ in the two half spaces, and this leads to similarly rotated quasiparticle wavevectors.  Thus, in the two superconducting half spaces, we take ${\bf k}_{\eta}=(k_{x\eta},k_{y\eta})$, which are rotated by $\pm\phi_0/2$ for $\eta=L,U$, respectively.  For any OP symmetry, we 
thus have $\xi_{0\eta}({\bf k})=\xi_0({\bf k}_{\eta})$ and $\Delta_{\eta}({\bf k})=\Delta({\bf k}_{\eta})$.  Although for an isotropic $s$-wave pairing interaction, the twist orientation doesn't matter as far as the OP $\Delta$ is concerned, a non-vanishing $\phi_0$ still causes the quasiparticle dispersions to be different in the two superconductors.   This leads to a strong impedance mismatch for all $\phi_0$ values not too close to $0^{\circ}, 90^{\circ}$.  

In Figs. 5-8, we have presented our results for $I_c(\phi_0)/I_c(0)$ at $T/T_c=0.5$ for different values of the material parameters.  In Fig. 5, we presented our  results for the optimally doped tight-binding quasiparticle dispersion, Eq. (\ref{dispersion}), in which  $\mu=0.6$, $\nu=1.3$, and $J_{||}=500$ meV, with Fermi surface FS2.  We include curves for each of the three OP symmetries, and for $J_{\eta}=25, 50, 100$ meV. For the extended-$s$-wave OP, we set $\epsilon=0$, so that the wavevector dependence of the OP has the $|\cos(k_xa)-\cos(k_ya)|$ form.  Results for the extended-$s$-wave OP with $\epsilon>0$ are intermediate to the extended-$s$ and $s$-wave results shown.

\begin{figure}[htb]
\vspace*{-0.0cm}
\epsfxsize=7.0cm
\centerline{\epsffile{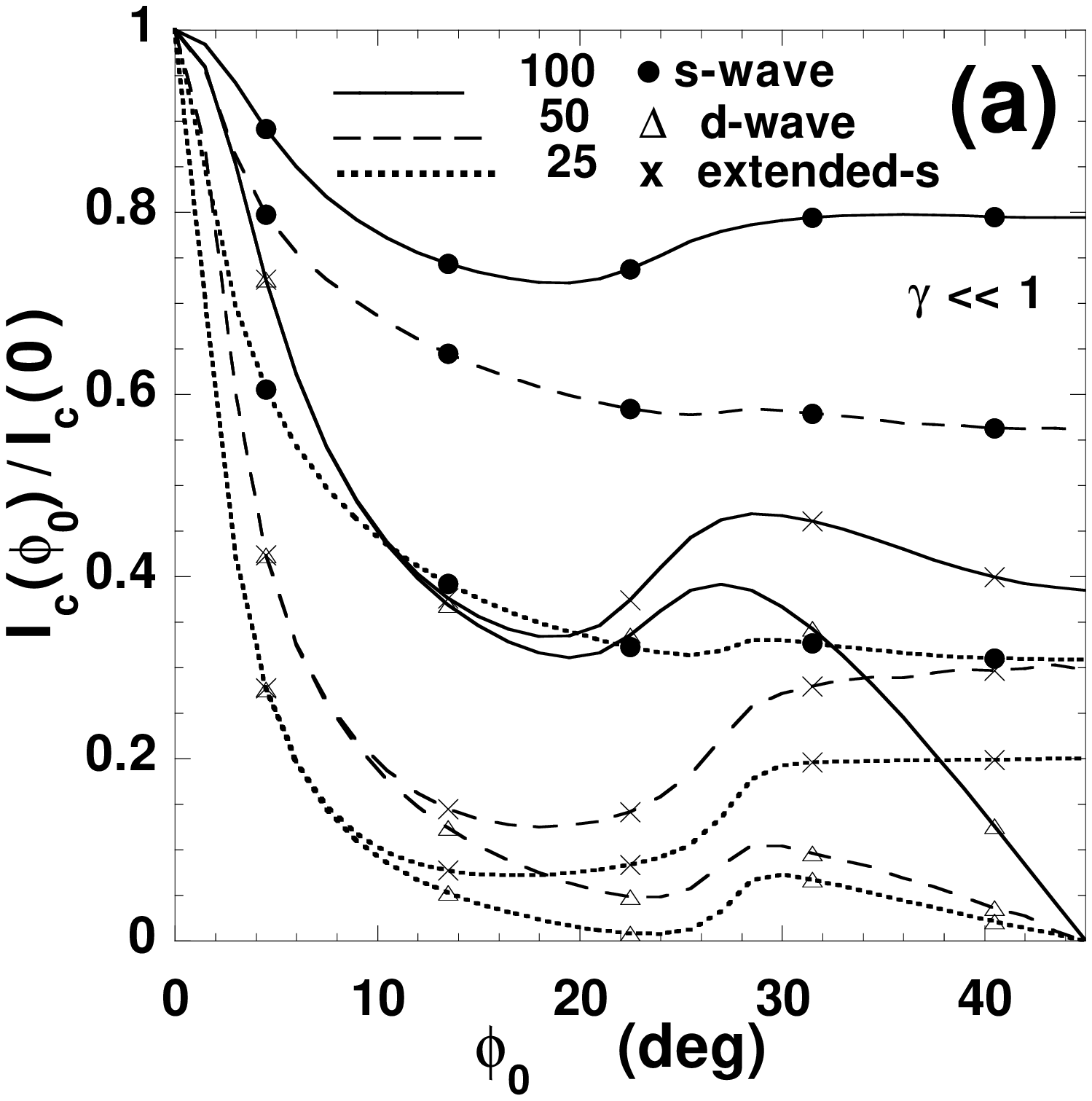}}
\end{figure}
\begin{figure}[htb]
\vspace*{-1.0cm}
\epsfxsize=7.0cm
\centerline{\epsffile{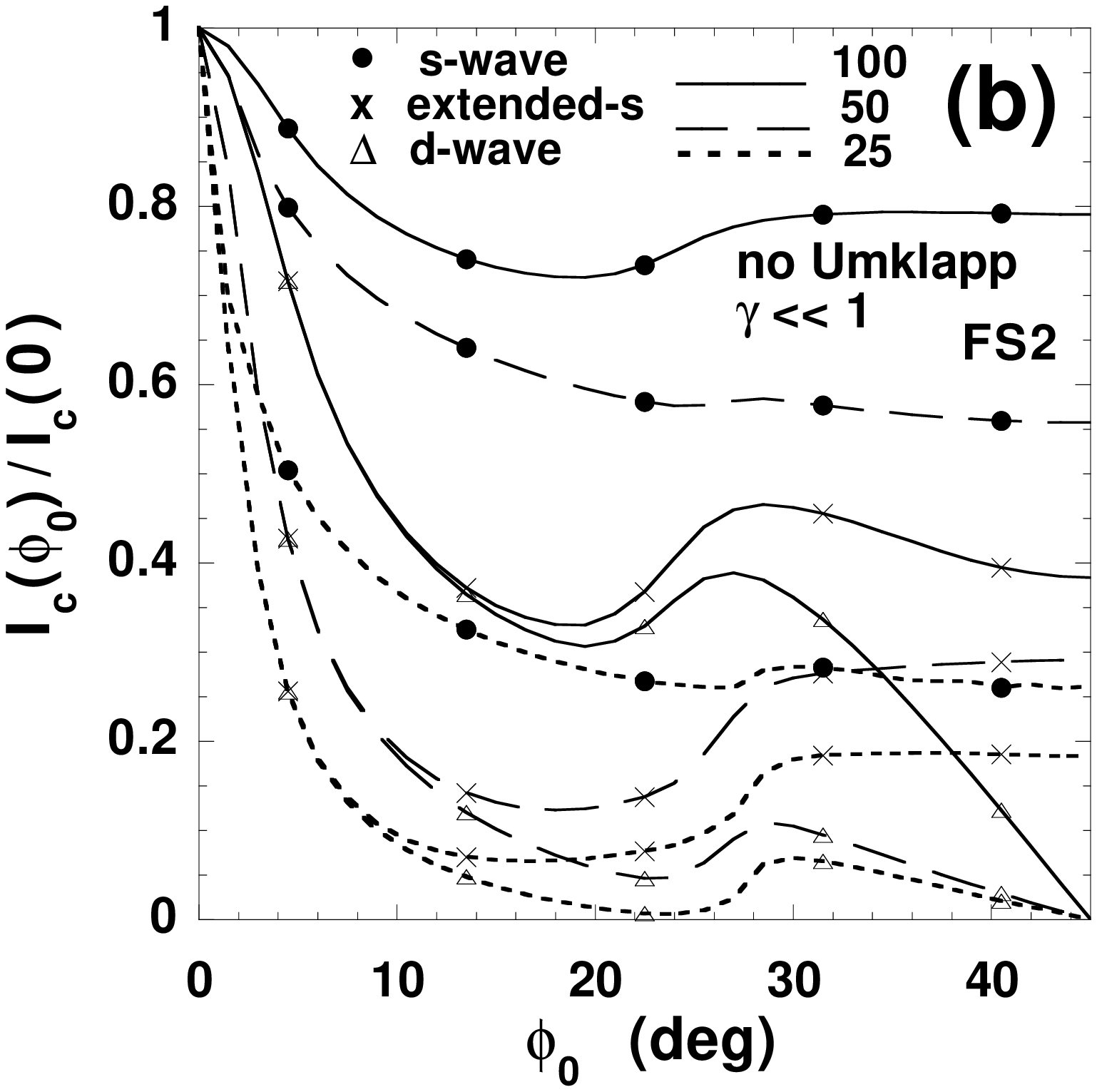}}
\end{figure}
\begin{figure}[htb]
\vspace*{-1.0cm}
\epsfxsize=7.0cm
\centerline{\epsffile{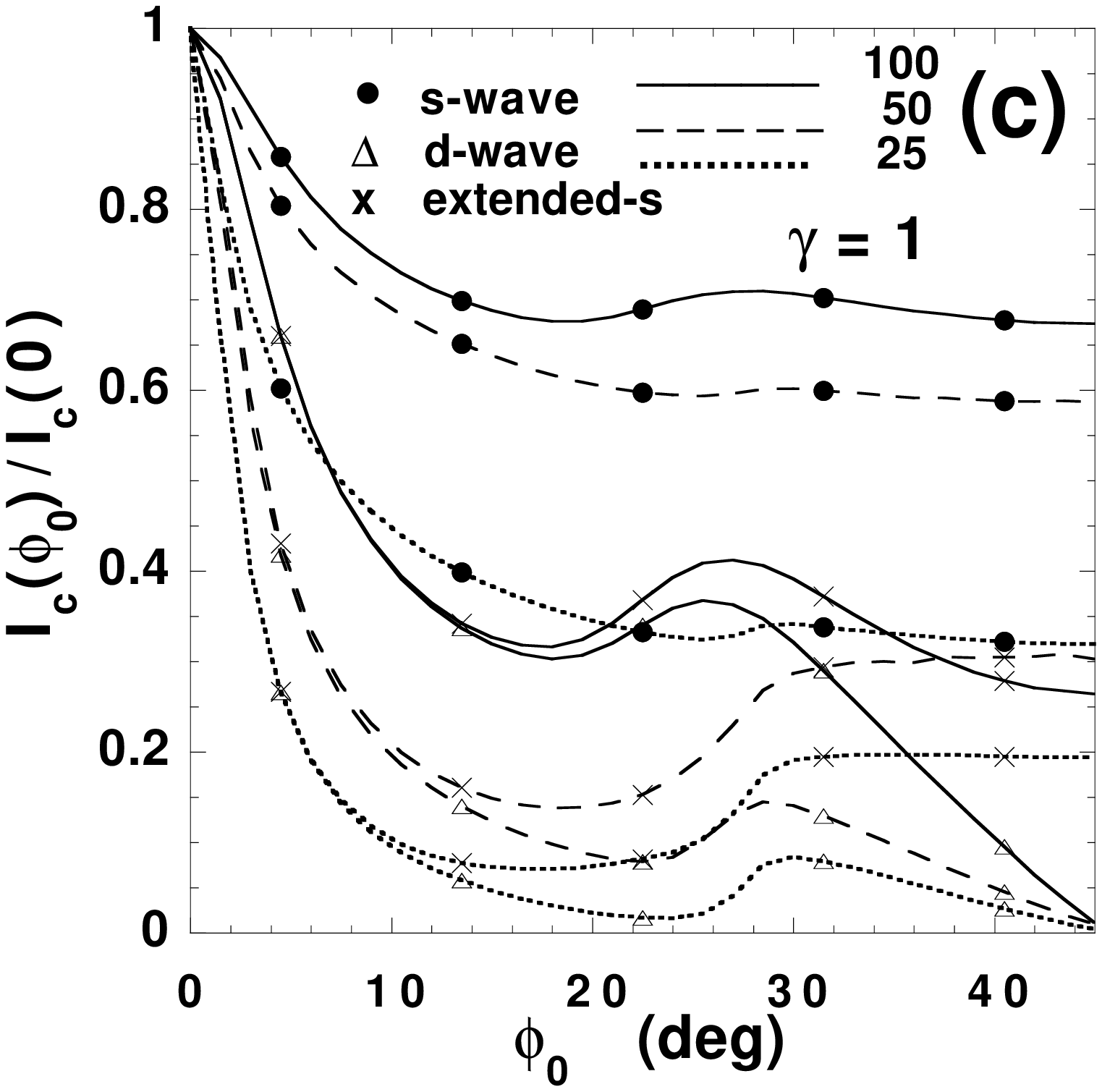}}
\vspace*{0.3cm}
\caption{Plots of $I_c(\phi_0)/I_c(0)$ at $T/T_c=0.5$ for $c$-axis twist junctions between layered superconductors with FS2 and OPs of $s$-, $d$-, and extended-$s$-wave symmetry, for $J_{\eta} = 25, 50, 100$ meV.  (a)  $\gamma\ll1$, including Umklapp processes.  (b) $\gamma\ll1$, without Umklapp processes.  (c)  $\gamma=1$ with Umklapp.}\label{Fig5}
\end{figure}

 In each case, there are both direct and Umklapp tunneling processes.  In the direct processes, a quasiparticle undergoing tunneling across the twist junction has wavevectors ${\bf k}_L$ and ${\bf k}_U$ both within the first BZ.  However, for the case in which either the initial or the final wavevectors is outside of the respective first BZ, Umklapp processes can occur, due to the periodic nature of the quasiparticle dispersions assumed.  To investigate whether these Umklapp processes are important, we treat the two limiting cases, either (1) that they can be completely neglected, or (2) that they are equal in weight to the direct processes. 

 In Fig. 5a, we set $\gamma\ll1$, the tunneling Hamiltonian limit.   In this case, we included the Umklapp processes with the same weighting as for direct tunneling processes within the first BZ on each side of the twist junction.    Preliminary versions of Fig. 5a with slightly different parameters were presented earlier. \cite{KBRSA}  For comparison, in Fig. 5b, we used the same parameters as in Fig. 5a, but the Umklapp processes were completely excluded.  We note that for this quasiparticle dispersion, not very much of the intersection of the Fermi surfaces is excluded by neglecting the Umklapp processes for a twisted junction.  In addition, in Fig. 5c, we presented our results for $\gamma=1$, including the Umklapp processes.  We note that comparing these results with those of Fig. 5b, the main differences occur for the larger $J_{\eta}$ values, especially $J_{\eta}=100$ meV.  Otherwise, for small $J_{\eta}$, there is very little difference between them.

In addition, we note that the main differences between the $d_{x^2-y^2}$-wave and extended-$s$-wave $I_c(\phi_0)$ results appear for $\phi_0$ close to 45$^{\circ}$. As $\phi_0\rightarrow45^{\circ}$, $I_c(\phi_0)/I_c(0)\rightarrow0$ for the $d$-wave OP, whereas for the extended-$s$-wave OP, $I_c(\phi_0)/I_c(0)$  remains finite and flattens out, becoming only weakly dependent on $\phi_0$.  Note that for Fermi surface FS2, even the isotropic $s$-wave OP leads to an anisotropic $I_c(\phi_0)/I_c(0)$, for each of the $J_{\eta}$ values shown, reflecting the impedance mismatch across the twist junction.  The greatest $\phi_0$ variation occurs for the smallest $J_{\eta}$ values, as the quasiparticle dispersions are the most two-dimensional, with the greatest impedance mismatch occurring for $\phi_0$ far from $0^{\circ},90^{\circ}$.

In Fig. 6, we again plotted $I_c(\phi_0)/I_c(0)$ at $T/T_c=0.5$ fro $c$-axis twist junctions between optimally doped sampels  with Fermi surface FS2 in the tunneling Hamiltonian limit $\gamma\ll1$, including the Umklapp processes.  But, we now used the wavevector-dependent coherent interlayer tunneling similar to that suggested by Liechtenstein {\it et al.} \cite{Andersen}  We note that this wavevector-dependent interlayer tunneling gives rise to a stronger $\phi_0$ dependence of $I_c$ than does the constant, wavevector-independent tunneling model. In addition, for this model, the extended-$s$ and $d$-wave OPs give rise to nearly identical $I_c(\phi_0)/I_c(0)$ curves, except for $\phi_0$ near to $45^{\circ}$, of course, where $I_c(\phi_0)\rightarrow0$ for the $d$-wave case, but not for the other two OPs. 

\begin{figure}[htb]
\vspace*{-0.0cm}
\epsfxsize=7.0cm
\centerline{\epsffile{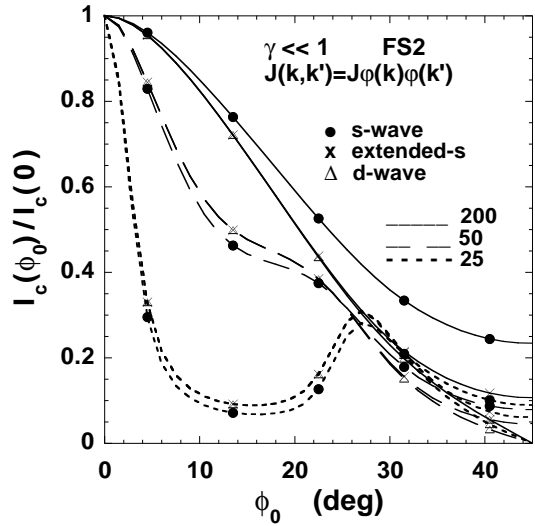}}
\vspace*{0.3cm}
\caption{Plots of $I_c(\phi_0)/I_c(0)$ at $T/T_c=0.5$ and $\gamma\ll1$ with Umklapp for $c$-axis twist junctions between layered superconductors with FS2 and $J({\bf k},{\bf k}')=J\varphi({\bf k})\varphi({\bf k}')$, $J_{\eta}({\bf k}_{\eta}, {\bf k}_{\eta})=J_{\eta}\varphi^2({\bf k}_{\eta})$, where $\varphi({\bf k})=|\cos(k_xa)-\cos(k_ya)|$, ${\bf k}_U={\bf k}$ and ${\bf k}_L={\bf k}'$, $J_{\eta} = 25, 50, 200$ meV and $s$-, $d$-, and extended-$s$-wave OP symmetry.}\label{Fig6}
\end{figure}

In Fig. 7, we plotted $I_c(\phi_0)/I_c(0)$ at $T/T_c=0.5$ in the tunneling Hamiltonian limit ($\gamma\ll1$) for $c$-axis twist junctions between  heavily underdoped samples with Fermi surface FS1, including Umklapp processes, and kept $J_{\eta}$ fixed at 100 meV.  Although curves are shown for each of the three OP symmetries studied, for this quasiparticle dispersion, it is difficult to distinguish them, as the $I_c(\phi_0)/I_c(0)$ values all either vanish (for the $d$-wave OP) or become very small for $\phi_0\approx45^{\circ}$.

\begin{figure}[htb]
\vspace*{-0.0cm}
\epsfxsize=7.0cm
\centerline{\epsffile{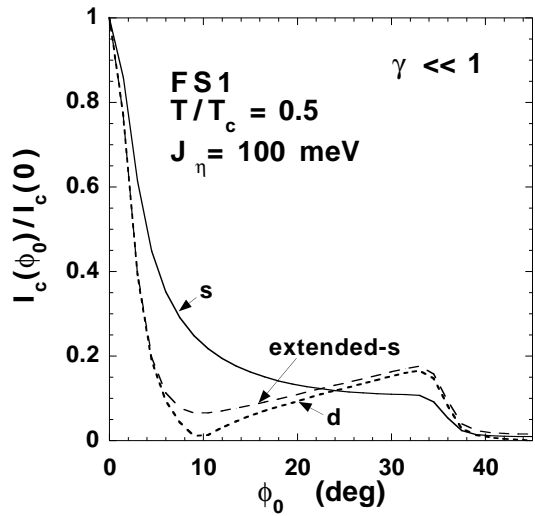}}
\vspace*{0.3cm}
\caption{Plots of $I_c(\phi_0)/I_c(0)$ at $T/T_c=0.5$ and $\gamma\ll1$ with Umklapp for $c$-axis twist junctions of layered superconductors  with FS1 with $J_{\eta}=100$ meV, and $s$-, $d$-, and extended-$s$-wave OP symmetry.}\label{Fig7}
\end{figure}

Finally, in Fig. 8, we presented plots of $I_c(\phi_0)/I_c(0)$ at $T/T_c=0.5$ in the tunneling Hamiltonian limit ($\gamma\ll1$) for $c$-axis twist junctions between heavily overdoped samples with Fermi surface FS3.  Results for $J_{\eta}=25,50,100$ meV are shown.  In this case, there is very little impedance mismatch for the isotropic $s$-wave OP, since the rotation induced by the twist junction nearly maps the Fermi surface onto itself. One of these curves (the $s$-wave OP curve with $J_{\eta}=100$ meV) is actually consistent with the data of Li {\it et al.}, although the quasiparticle dispersion assumed is very metallic and three-dimensional, completely inconsistent with that expected for BSCCO.  The $s$-wave curve with $J_{\eta}=50$ meV is only  marginally consistent with the data at best, and the $s$-wave curve with smaller $J_{\eta}$ values can be ruled out.  Similarly, the extended-$s$ and especially the $d$-wave curves are all inconsistent with the experimental data.

\begin{figure}[htb]
\vspace*{-0.0cm}
\epsfxsize=7.0cm
\centerline{\epsffile{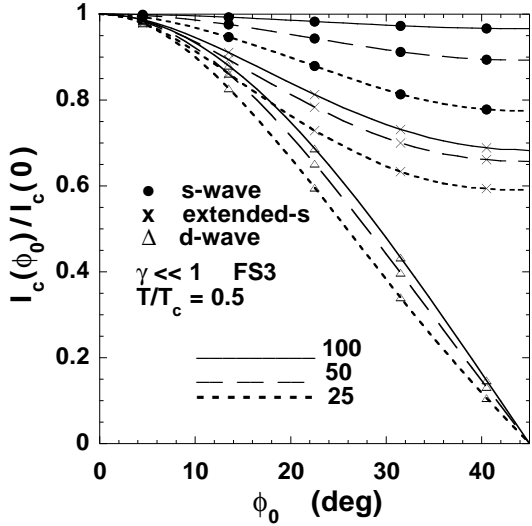}}
\vspace*{0.3cm}
\caption{Plots of $I_c(\phi_0)/I_c(0)$ at $T/T_c=0.5$ and $\gamma\ll1$ with Umklapp for $c$-axis twist junctions of layered superconductors with FS3 and  with $s$-, $d$-, and extended-$s$-wave OP symmetry.}\label{Fig8}
\end{figure}

\section{Conclusions}
We calculated the Josephson current for $c$-axis coherent tunneling between two layered superconductors exactly.  We assumed the intralayer quasiparticle dispersion of the tight-binding form, Eq. (\ref{dispersion}).  Since the parameters in this model are arbitrary, our results also apply to the cases in which one or both of the superconductors is a conventional, bulk material.  In these cases, one simply lets one or both of the $J_{\eta}\rightarrow    J_{\eta||}\rightarrow\infty$.  For $\gamma\ne1$, the tunneling properties across the junction are different than the intrinsic tunneling between adjacent layers far from the junction, and the role of surface states on the sides of the junction becomes important. 

Such surface states were first investigated for superconductors in the case of a vacuum interface. \cite{LK}  Then, the role of surface states upon the coherent Josephson tunneling between layered superconductors was investigated within the tunneling Hamiltonian limit. \cite{KARS}  Our present results comprise the first calculation of the strong coherent Josephson tunneling between two superconductors which properly takes account of the surfaces states. The earlier results of Tanaka and Kashiwaya, \cite{TK} which were claimed to give the correct results for this problem, are completely incorrect.  In those calculations, the quasiparticle dispersions were not of the proper layered, tight-binding form, and the approximations used were not correct in any limit. 

Our results also apply to the case of tunneling between two superconductors in other orientations.  For instance, in the case of an isotropic bulk superconductor, the surface states present along the sides of a junction normal to the $c$-axis are precisely the same as that along the sides of a junction in another orientation. Thus, the calculations done by others, including Tanaka and Kashiwaya, \cite{TK} for the Josephson tunneling within the $ab$-plane of layered superconductors, do not correctly take these surface states into account.  In addition, since they always assume a circular Fermi surface cross-section (similar to FS3 in Fig. 2), all effects of  impedance mismatching present in our investigation of the $c$-axis twist junctions are neglected.  As shown in Section VI, such effects can be very strong, especially when the OP is highly anisotropic.

Although we have not introduced the effects of incoherent $c$-axis tunneling, we found that coherent tunneling between two superconductors that are not precisely the same gives rise to strong impedance mismatch effects. For coherent tunneling between identical layered superconductors twisted about the $c$-axis with respect to each other, the impedance mismatch effects are strong. Even when the OP is isotropic, one expects a strong twist angle $\phi_0$ dependence of the Josephson critical current.  Although Umklapp processes are present in tunneling between superconductors twisted about the $c$-axis, they make only a very small correction to the actual critical current.

We calculated the Josephson critical current across $c$-axis twist junctions, for OPs  of the isotropic $s$-wave, the tight-binding $d_{x^2-y^2}$-wave, $\cos(k_xa)-\cos(k_ya)$, and the tight-binding `extended-$s$-wave', $\{[\cos(k_xa)-\cos(k_ya)]^2+\epsilon^2\}^{1/2}$, forms.  We studied the cases of relative weak tunneling ($\gamma\ll1$) and strong tunneling ($\gamma=1$), for various $c$-axis dispersion bandwidths.  Except for the simple case of an isotropic OP on a circular Fermi surface (akin to FS3), we conclude that coherent tunneling cannot possibly explain the data of Li {\it et al.}. \cite{Li} Moreover, since the $d_{x^2-y^2}$-wave OP cannot possibly fit the data of Li {\it et al.} under any circumstances, we conclude strongly that the OP is not $d$-wave near to $T_c$.  Since there is widespread agreement that the Fermi surface of BSCCO is unlikely to have a circular cross-section, our calculations thus support the contention of Li {\it et al.} that the $c$-axis tunneling intrinsic in BSCCO must be largely incoherent. \cite{Li}  A brief discussion of the role of incoherent tunneling has been presented, supporting this argument. \cite{KBRSA} A more thorough treatment of the role of incoherent tunneling will be discussed in detail elsewhere. \cite{BSK}

\section{acknowledgments}
The authors would like to thank A. Bille, Qiang Li, and K. Scharnberg for useful discussions.  This work was supported by the USDOE-BES through Contract No. W-31-109-ENG-38. 

\section{Appendix:  Exact Solutions for Green's Functions in the Layers}
To obtain the elements
${\cal G}_{mn}$ of the bulk tight-binding Green's function matrix $\hat{\cal G}$,
we can first Fourier series transform in the layer index, letting
\begin{equation}
{\cal G}_{mn}^{\eta}({\bf k}, \omega)=\int_{-\pi/s}^{\pi/s}{{sdk_z}\over{2\pi}}\exp[ik_zs(m-n)]{\cal G}^{\eta}({\bf k},k_z,\omega).\label{FT}
\end{equation} 
Eq. (\ref{bulkmatrix}) can then be solved algebraically to give
\begin{equation}
{\cal G}^{\eta}({\bf k},k_z,\omega)=-{{i\omega+\Delta_{\eta}\tau_1+\xi_{\eta}\tau_3}\over{\omega^2+|\Delta_{\eta}|^2+\xi_{\eta}^2}},
\end{equation}
where the quasiparticle dispersion
\begin{equation}
\xi_{\eta}({\bf k},k_z)=\xi_{0\eta}({\bf k})-J_{\eta}\cos k_zs.
\end{equation}
For notational simplicity
in the following, we shall omit the ${\bf k}$ and $\omega$
dependencies of the various functions.

Performing the integral in Eq. (\ref{FT}) and multiplying by $(J_{\eta}/2)\tau_3$ on the right, we write
\begin{eqnarray}
{\cal G}_{mn}^{\eta}{{J_{\eta}}\over{2}}\tau_3&=&\beta_{mn}^{\eta}+(\epsilon_{\eta}\tau_3-\delta_{\eta}\tau_2)\alpha_{mn}^{\eta}\cr
& &\cr
&\equiv&
\beta_{mn}^{\eta}+\vec\alpha_{mn}^{\eta}\cdot\vec\tau,
\label{bulkgf}\end{eqnarray}
where
\begin{equation} \alpha_{mn}^{\eta}= (\phi_{mn}^{\eta +} 
-\phi_{mn}^{\eta -})/2,\end{equation}
\begin{equation} \beta_{mn}^{\eta}=(\phi_{mn}^{\eta +}
 +\phi_{mn}^{\eta -})/2,\end{equation}
and
\begin{equation}
\phi_{mn}^{\eta\pm}={{\pm i\exp(\pm i\theta_{\eta\pm}|
n-m|)}\over{2\sin(\theta_{\eta\pm})}},
\end{equation}
with
\begin{equation}
\cos(\theta_{\eta\pm})={{\xi_{0\eta}\pm i\Omega_{\eta}}\over{J_{\eta}}}.
\end{equation}
The parameter $\theta_{\eta +}$ is defined to have a positive imaginary part, and
$\theta_{\eta -}$ is defined to have negative imaginary part.
For convenience, we have defined
$\Omega_{\eta}=\sqrt{\omega^2+|\Delta_{\eta}|^2}$,
$\epsilon_{\eta}=\omega/\Omega_{\eta}$, and $\delta_{\eta}=\Delta_{\eta}/\Omega_{\eta}$.
For comparison with previous notation, \cite{KARS} we then write
\begin{equation}
\exp(i\theta_{\eta+})={{J_{\eta}}\over{2}}\Xi_{\eta},
\end{equation}
where $\Xi_{\eta}$ is given by Eq. (\ref{xi}).

We then find the half-space Green's functions, using the procedure of Eqs. (\ref{veeeta}) and (\ref{geeeta}). 
For the $L$ half space defined by $m,n\le 0$, we have
\begin{equation}
g_{mn}^L={\cal G}_{mn}^L-{\cal G}_{m1}^L{J_L\over2}\tau_3g_{0n}^L,
\end{equation}
which is easily solved for $g_{0n}^L$ by setting $m=0$, yielding
\begin{equation}
g_{mn}^L={\cal G}_{mn}^L-({\cal G}_{m1}^L{J_L\over2}\tau_3)[1+{\cal G}_{01}^L{J_L\over2}\tau_3]^{-1}
{\cal G}_{0n}^L.
\end{equation}
Similarly, for the $U$ half space defined by $m,n\ge 1$, we find
\begin{equation}
g_{mn}^U={\cal G}_{mn}^U-({\cal G}_{m0}^U{J_U\over2}\tau_3)[1+{\cal G}_{10}^U{J_U\over2}\tau_3]^{-1}
{\cal G}_{1n}^U.
\end{equation}

After some straightforward algebra, one finds for each half space
\begin{eqnarray}
g_{mn}^{\eta}{J_{\eta}\over2}\tau_3&\equiv &\bar{g}_{mn}^{\eta}=
b_{mn}^{\eta}+(\epsilon_{\eta}\tau_3-\delta_{\eta}\tau_2)a_{mn}^{\eta}\cr
& &\cr
& &\equiv b_{mn}^{\eta}+\vec{a}_{mn}^{\eta}\cdot
\vec{\tau},
\end{eqnarray}
where
\begin{equation}
a_{mn}^{\eta}={1\over2}(\chi_{mn}^{\eta +}-\chi_{mn}^{\eta -})\label{amn}
\end{equation}
and
\begin{equation}
b_{mn}^{\eta}={1\over2}(\chi_{mn}^{\eta +}+\chi_{mn}^{\eta -})\label{bmn}
\end{equation}
for $\eta=L,U$. For $n\le m\le 0$, we have
\begin{equation}
\chi_{mn}^{L\pm}={{-\exp[\mp
i\theta_{L\pm}(n-1)]\sin[\theta_{L\pm}(m-1)]}\over{\sin(\theta_{L\pm})}}.
\end{equation}
For $1\le m\le n$,
\begin{equation}
\chi_{mn}^{U\pm}={{\exp(\pm
i\theta_{U\pm}n)\sin(\theta_{U\pm}m)}\over
{\sin(\theta_{U\pm})}}.\end{equation}

To allow for different OP phases in the two sides of the
junction, we introduce a phase factor multiplying the order parameter
in the left hand side, letting 
\begin{equation}
\delta_L=|\delta_L|\exp[i(\phi_L-\phi_U)].
\end{equation}
The function $\delta_U$ thus does not include such a phase factor.

So far, we have found expressions for the two distinct half space Green's functions, which are electronically uncoupled from each other, since no quasiparticle propagation from one half space to the other has yet been introduced. We thus couple them together via the local
perturbation $\hat{\cal J}$ with matrix elements given by Eq. (\ref{coupling}).

The exact solution to this problem of coupled half spaces then yields the full Green's function matrix $\hat{G}$, with matrix elements $G_{mn}$.   In the $L$  half
space, $n,m\le 0$, the $G_{mn}$ satisfy
\begin{equation} G_{mn}{{J_L}\over{2}}\tau_3=\bar{g}_{mn}^L+\gamma
\bar{g}_{m0}^L[(\bar{g}_{11}^U)^{-1} -\gamma
\bar{g}_{00}^L]^{-1}
\bar{g}_{0n}^L.\label{exactGL}\end{equation}

Now the gap equation, Eq. (\ref{gapeqn}) in Section III can be found from $G_{nn}$.  On the $L$ side of the junction, we then need $G_{00}$. From Eq. (\ref{exactGL}) we find that the exact Green's function at the
interface and in the lower superconductor is
\begin{equation}
G_{00}{J_L\over2}\tau_3={{\bar
{g}_{00}^L-\gamma|\exp(i\theta_{U+}+i\theta_{L+})|^2 (\bar
{g}_{11}^U)^{-1}}\over{D}},\label{G00J3}\end{equation} where
\begin{eqnarray}
D&=&1+\gamma^2\vert\exp{(i\theta_{U+}+i\theta_{L+})}\vert^2\cr
& &\cr
& &-2\gamma
\{[\epsilon_L
\epsilon_U+\delta_U\delta_L\cos(\phi_L-\phi_U)]
A_UA_L\cr
& &\cr
& &\quad+B_UB_L\},
\end{eqnarray} 
\begin{equation} 
A_{\eta}=[\exp(i\theta_{\eta +})-\exp(-i\theta_{\eta -})]/2\end{equation}
and \begin{equation} 
B_{\eta}=[\exp(i\theta_{\eta +})+\exp(-i\theta_{\eta -})]/2.\end{equation}

Then, the trace in Eq. (\ref{gapeqn}) can be evaluated by first multiplying Eq. (\ref{G00J3}) by $2\tau_3/J_L$ on the right side,
Leading to
\begin{eqnarray}
 {\rm Tr}[(\tau_1+i\tau_2)G_{00}]&=&{{-2i}\over{J_LD}}[A_L\delta_L\cr
& &\cr
& &+\gamma|\exp(i\theta_{L+})|^2A_U\delta_U].
\end{eqnarray}

This then leads to Eq. (9) in the text. 

In order to calculate the Josephson tunneling current, we need to find $G_{01}$ and $G_{10}$ explicitly. Solving for these functions yields the exact results
\begin{equation} G_{01}{J\over2}\tau_3=\gamma \bar{g}_{00}^L
[(\bar{g}_{11}^U)^{-1}-\gamma
\bar{g}_{00}^L]^{-1}\end{equation}
and
\begin{equation} G_{10}{J\over2}\tau_3=\gamma
[(\bar{g}_{11}^U)^{-1}-\gamma \bar{g}_{00}^L]^{-1}
\bar{g}_{00}^L.\end{equation}
Combining, and evaluating the trace, we find
\begin{equation} I=4e\gamma T\sum_{\bf k}\sum_{\omega}
{{|\delta_L||\delta_U| A_U A_L}\over{D}}
\sin(\phi_U-\phi_L).\label{Josephson}\end{equation}  From this expression, it is elementary to obtain Eq. (\ref{exactcurrent}) of the text.

In the tunneling Hamiltonian limit,  $\gamma<<1$, this reduces to
\begin{eqnarray} 
I&=&-4e\gamma T\sum_{\bf k}\sum_{\omega}|\delta_L|
|\delta_U| \Im[\exp(i\theta_{U+})]\times\cr
& &\cr
& &\quad\times\Im[\exp(i\theta_{L+})]
\sin(\phi_U-\phi_L).\end{eqnarray}
This result agrees with the tunneling Hamiltonian result for the coherent tunneling limit, as expected. \cite{KARS}


\begin{thebibliography}{10}
\bibitem{TK}
Y. Tanaka and S. Kashiwaya, Phys. Rev. B {\bf 56}, 892 (1997).
\bibitem{KS}
D. Kalkstein and P. Soven, Surf. Sci. {\bf 26}, 85 (1971).
\bibitem{LK}
S. H. Liu and R. A. Klemm, Phys. Rev. B {\bf 52}, 9657 (1995).
\bibitem{KARS}
R. A. Klemm, G. Arnold, C. T. Rieck, and K. Scharnberg, Phys. Rev. B {\bf 58}, 14 203 (1998). 
\bibitem{AB}
V. Ambegaokar and A. Baratoff, Phys. Rev. Lett {\bf 10}, 486 (1963); {\bf 11}, 104 (1963).
\bibitem{MK}
M. M{\"o\ss}le and R. Kleiner, Phys. Rev. B {\bf 59}, 4486 (1999).
\bibitem{Andersen}
A. I. Liechtenstein, O. Gunnarsson, O. K. Andersen, and R. M. Martin, 
Phys. Rev. B {\bf 54}, 12 505 (1996).
\bibitem{Li}
Q. Li, Y. N. Tsay, M. Suenaga, R. A. Klemm, G. D. Gu, and N. Koshizuka, Phys. Rev. Lett. {\bf 83}, 4160 (1999).
\bibitem{KRS}
R. A. Klemm, C. T. Rieck, and K. Scharnberg, Phys. Rev. B {\bf 58}, 1051 (1998);
{\it ibid.} {\bf 61}, to be published.
\bibitem{KBRSA}
R. A. Klemm, A. Bille, C. T. Rieck, K. Scharnberg, and G. Arnold, J. Low Temp. Phys. {\bf 117}, 509 (1999);  R. A. Klemm, G. Arnold, A. Bille, C. T. Rieck, and K. Scharnberg, Int. J. Mod. Phys. B {\bf 13}, 3449 (1999).  There was unfortunately an error in the program that evaluated Fig. 2b of the former paper, which we did not discover until after the manuscript was sent to press.  We were able to make the correction in Fig. 3b of the second paper, which shows the correct results for the same set of parameters.  The main differences appear in the $s$-wave and extended-$s$-wave curves, for which the correct curves exhibit much stronger $\phi_0$ dependencies of $I_c(\phi_0)/I_c(0)$ than do the incorrect curves.  Similar correct curves are shown here in Fig. 5 for a slightly different set of parameters.
\bibitem{BSK}
A. Bille, K. Scharnberg, and R. A. Klemm, unpublished.
\end{thebibliography}
\end{document}